\documentclass[aps,pra,reprint,amsmath,amssymb,showkeys,notitlepage]{revtex4-2}
\usepackage{color}
\usepackage{braket}
\usepackage{comment}
\usepackage{graphicx} 
\usepackage{txfonts}  
\usepackage{epstopdf}
\usepackage{appendix}

\usepackage{stackengine}
\usepackage{soul,xcolor}
\usepackage{mathtools}
\usepackage{bbold}
\usepackage{braket}
\usepackage{comment}
\newcommand{\md}{\mathrm{d}}
\begin{document}

\title{Strongly entangled system-reservoir dynamics with multi-photon pulses beyond the two-excitation limit: \\ Exciting the atom-photon bound state}

\author{Kisa Barkemeyer}
\thanks{k.barkemeyer@tu-berlin.de}
\author{Andreas Knorr}
\author{Alexander Carmele}
\affiliation{Institut f\"ur Theoretische Physik, Technische Universit\"at Berlin, 10623 Berlin, Germany}

\begin{abstract}

 Within the matrix product state framework, we study the non-Markovian feedback dynamics of a two-level system interacting with the electromagnetic field inside a semi-infinite waveguide where the excitation of an atom-photon bound state is possible.
Taking the steady-state excitation of the emitter as a figure of merit, we compare the trapped excitation for an initially excited quantum emitter and an emitter prepared via quantized pulses containing up to four photons. In the latter case, we find that for large feedback delay times, multi-photon pulses can yield a significantly higher steady-state excitation than possible with an initially excited emitter since the stimulated emission process can enhance the trapping probability in comparison to the spontaneous decay of an initially excited emitter.
\end{abstract}

\maketitle

\section{Introduction}\label{sec:introduction}

A promising platform for a reliable implementation of large-scale quantum networks
is offered by photonic quantum technologies where photons transport quantum information between the nodes of the network. The nodes, in turn, allow the storage as well as the manipulation of the information \cite{Cirac1997a,Parkins1999,DiVincenzo2000, Knill2001a, Zoller2005, Kimble2008a,OBrien2009,Nielsen2010,Northup2014,Vermersch2017a}.
In recent years, special attention has been paid to waveguide quantum electrodynamics (w-QED) systems consisting of quantum few-level systems interacting with the electromagnetic field inside a one-dimensional waveguide. In these systems, enhanced light-matter interaction and interference effects can be observed due to the spatial confinement of the light field \cite{Chang2006,Fan2010}. This way, they allow for the
creation of strong effective photon-photon interactions and qubit-qubit entanglement and, thus, are eligible candidates for the realization of quantum information processing protocols \cite{Chang2007,Ciccarello2012, Zheng2013,Zheng2013b,Gonzalez2014}. 

Such w-QED systems have been studied extensively in the Markovian regime where employed methods include the input-output formalism \cite{Fan2010,Kiilerich2019,Kiilerich2020,Liao2020},
the Lippmann-Schwinger equation \cite{Shen2007,Zheng2013,Fang2014},
a Green's function approach \cite{Dzsotjan2010,Schneider2016},
as well as master equations \cite{Baragiola2012,Chang2012}.
The Markovian approximation, however, breaks down if there is a macroscopical separation between the nodes compared to the wavelength of the light, for example, in long-distance networks. In this case, non-Markovian effects become important since the time-delayed backaction of the electromagnetic field on the emitters has to be taken into account.
As a consequence, the possibility to use time-delayed signals, for example, in the context of sub- and superradiance or coherent feedback control, is opened up
\cite{Wiseman1994,Lloyd2000,Dorner2002,Carmele2013,Hein2014,Tufarelli2014,Grimsmo2015,Kabuss2015c,Kabuss2016,Chalabi2018,Fang2018,Nemet2019,Barkemeyer2019,Sinha2020,Carmele2020}. 
Methods to deal with non-Markovian system dynamics include scattering theory \cite{Fang2015,Guimond2017} and non-Markovian quantum state diffusion \cite{Diosi1998}. A further method to treat such systems is the matrix product state (MPS) framework \cite{Schollwock2011,Pichler2016,Guimond2017,Carmele2020}.

A remarkable feature in w-QED systems is the formation of atom-photon bound states. On the one hand, the interaction of the light field with quantum impurities in finite-bandwidth waveguides can result in bound states outside the continuum of propagating modes \cite{Longo2010,Calajo2016}. On the other hand, using for example time-delayed feedback, it is also possible to excite bound states inside the continuum which can potentially be used for the storage of quantum information \cite{Longhi2007,Lvovsky2009, Redchenko2014,Facchi2016,Hsu2016,Saglamyurek2018,Dinc2019,Facchi2019,Finsterholzl2020,Leonforte2020}.
A paradigmatic setup in this regard is the two-level system (TLS) in front of a mirror where emission properties depend sensitively on the emitter-mirror separation.
Here, a finite excitation of the emitter in the long-time limit is possible due to the excitation of an atom-photon bound state. In this bound state, the excitation is distributed between the TLS (excitonic component) and the waveguide between the TLS and the mirror (photonic component). For small separations, the excitonic component dominates while for larger delays the photonic component becomes more important. There are two possibilities to populate the bound state: By letting an initially excited emitter decay or via multi-photon pulses. The first possibility is most efficient for small separations where a large overlap of the initial state of the system with the excitonic component of the bound state can be found and it holds that the smaller the separation, the higher the steady-state excitation of the emitter \cite{Tufarelli2013,Hoi2015a}. If the bound state is addressed via photon scattering, the steady-state excitation depends non-monotonously on the emitter-mirror separation. The effectiveness of this excitation scheme is determined by the overlap of the initial state of the system with the photonic component of the bound state. Thus, a certain minimum separation is crucial for it to work.
Because of this non-monotonous behavior, studies with pulses of various shapes containing variable numbers of photons are of interest for which an efficient description needs to be developed \cite{Calajo2019a,Cotrufo2019,Trivedi2020}.

Here, we study the excitation of an atom-photon bound state for a TLS inside a semi-infinite one-dimensional waveguide using the MPS framework for pulses containing up to four photons. 
Thereby, we extend an existing method for pulses containing up to two photons \cite{Guimond2017}. In this context, we look for ways to control and, in particular, maximize the steady-state excitation of the emitter.

The Paper is structured as follows: After this introduction in Sec.~\ref{sec:introduction}, in Sec.~\ref{MPS_method}, we introduce the considered system and present
the MPS method for the calculation of its dynamics including quantized pulses which we benchmark using an approach in the Heisenberg picture. In Sec.~\ref{MPS_results}, we discuss the non-Markovian system dynamics and, in particular, the excitation of the atom-photon bound state. Finally, we summarize our findings in Sec.~\ref{Conclusion}.

\section{MPS method}
\label{MPS_method}

\begin{figure}[htb]
    \centering
    \includegraphics[width=0.7\linewidth]{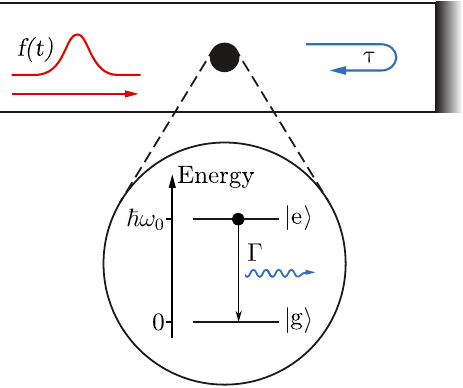}
    \caption{(Color online) TLS with decay rate $\Gamma$ consisting of ground state $\ket{g}$ and excited state $\ket{e}$ separated by energy $\hbar \omega_0$ inside a semi-infinite one-dimensional waveguide which provides feedback at the delay time $\tau$. The TLS is excited via a quantum pulse of shape $f(t)$.}
    \label{fig:TLS_Setup}
\end{figure}

Here, we present a method in the MPS framework that allows the numerically exact calculation of the dynamics in w-QED systems \cite{Schollwock2011,Pichler2016,Guimond2017}. We use the approach to study the dynamics of a single TLS inside a semi-infinite waveguide. The closed end of the waveguide at a distance $d$ from the TLS functions as a mirror. It feeds back the excitation emitted from the TLS after a delay time $\tau = 2d/c$ where $c$ is the speed of light in the waveguide. The combined system of the TLS and the photonic reservoir is depicted in Fig.~\ref{fig:TLS_Setup} and can be described via the Hamiltonian in dipole and rotating wave approximation
\begin{align}
\mathcal{H} &= \mathcal{H}_0 + \mathcal{H}_\text{int},\\
\mathcal{H}_0 &= \hbar \omega_0 E +  \hbar \int d\omega \omega r_\omega^\dagger r_\omega, \\ \mathcal{H}_{\text{int}} &= \hbar \int d\omega g(\omega) \left( r_\omega^\dagger \sigma_- + \text{H.c.}\right)
\end{align}
where $\omega_0$ is the transition frequency of the TLS and the operator $E=\sigma_+\sigma_-$ describes the occupation of its excited state. Its lowering (raising) operator is denoted by $\sigma_-$ ($\sigma_+$) which, thus, can be interpreted as the polarization of the TLS and it holds that $\sigma_+ = \sigma_-^\dagger$. If we assume a TLS containing at most one excitation, we find $\left[ \sigma_-,\sigma_+\right] = \mathbb{1}-2E$. The annihilation (creation) of a photon with frequency $\omega$ in the reservoir is described by the bosonic operator $r_\omega^{(\dagger)}$. Reservoir and TLS are coupled with strength $g(\omega)$ which is, in general, frequency-dependent.
Since we want to model feedback effects we consider a structured reservoir which results in the frequency-dependent coupling strength $g(\omega) = g_0 \sin(\omega \tau /2)$ where $\tau$ is the feedback-induced delay time.

We transform the Hamiltonian $\mathcal{H}$ into the rotating frame defined by its non-interacting part $\mathcal{H}_0$ which yields
\begin{equation}
    \mathcal{H}' = \hbar \int d\omega  g(\omega) \left( e^{i \left(\omega -\omega_0\right)t} r_\omega^\dagger \sigma_-  + \text{H.c.}\right). 
    \label{eq:H_int}
\end{equation}
The Hamiltonian $\mathcal{H}'$ governs the time evolution of the system which we discuss in the following section.

\subsection{Time evolution}
\label{TimeEvolution}

In this section, the time evolution method based on MPS is introduced in a condensed form to pave the way for the inclusion of quantized pulses.
For a detailed derivation of the time evolution algorithm with feedback see Ref.~\cite{Pichler2016}.
The MPS framework is based on the Schrödinger picture.
The main idea of the MPS time evolution method is to describe the continuous system dynamics via a stroboscopic time evolution at discrete time steps $\Delta t$ which are small compared to the time scales of the system evolution.
To start with, we introduce the time-dependent quantum noise operators \cite{Loudon}
\begin{equation}
    r^\dagger_t = \frac{1}{\sqrt{2\pi}}\int \md \omega r_\omega^\dagger e^{i\left(\omega-\omega_0\right)t}.
    \label{eq:noise_op}
\end{equation}
As the conjugate operators of $r_\omega^\dagger$, these collective operators describe the creation of a photon at time $t$ and satisfy $\left[r_t, r^\dagger_{t'} \right] = \delta(t-t')$. With this definition, after a time-independent phase shift via a unitary transformation \cite{Pichler2016,Carmele2020}, the Hamiltonian $\mathcal{H}'$ from Eq.~\eqref{eq:H_int} can be written as
\begin{equation}
    \mathcal{H}'(t) = - i \hbar \sqrt{\Gamma} \left[r^\dagger_t \sigma_- - r^\dagger_{t-\tau}\sigma_-e^{-i\omega_0 \tau} - \text{H.c.} \right]
    \label{eq:H_timedep}
\end{equation}
which is now explicitly time-dependent. Here, we defined the decay rate $\Gamma \equiv \pi g_0^2/2$. The first term on the right-hand side of Eq.~\eqref{eq:H_timedep} describes the immediate interaction of the TLS and the reservoir while the second term arises due to the feedback signal the effect of which is determined by the feedback phase $\phi \equiv \omega_0 \tau$. Here, we particularly focus on the special case of $\omega_0 \tau = 2\pi n$, $n \in \mathbb{N}$, as discussed in detail in Sec.~\ref{MPS_results}.

The dynamics of the system is governed by the Schr\"odinger equation
\begin{equation}
    \frac{\md}{\md t} \ket{\psi(t)} = -\frac{i}{\hbar} \mathcal{H}'(t)\ket{\psi(t)}.
\end{equation}
If we discretize time in sufficiently small steps $\Delta t$, the evolution from time $t_k$ to $t_{k+1}$, $t_k = k\Delta t$, $k \in \mathbb{N}$, can be described via the coarse-grained stroboscopic time evolution operator $U_k$ for which
\begin{align}
    \ket{\psi(t_{k+1}} &= U_k \ket{\psi(t_k)}, \\ U_k &= \exp\left[-\frac{i}{\hbar}\int_{t_k}^{t_{k+1}} \md t' \mathcal{H}'(t')\right].
\end{align}
Concretly, in our system, assuming $\tau = l \Delta t$, $l \in \mathbb{N}$, it takes the form
\begin{multline}
    U_k = \exp \left[ -\sqrt{\Gamma} \left[ \Delta R^\dagger (t_k)\sigma_- - \Delta R^\dagger(t_{k-l}) \sigma_-e^{-i\omega_0 \tau} \right.\right.\\ \left.\left.- \text{H.c.}  \vphantom{\Delta R^\dagger}\right]\vphantom{\sqrt{\Gamma}}\right]
\end{multline}
where we defined the noise increments
\begin{equation}
    \Delta R^\dagger(t_k) = \int_{t_k}^{t_{k+1}}\md t r^\dagger_t.
\end{equation}
These operators describe the creation of a photon in time step $k$ and obey $\left[ \Delta R(t_k),\Delta R^\dagger(t_{k'}) \right]= \Delta t \delta_{kk'}$.
With the noise increments, a discrete, orthonormal time-bin basis of the Hilbert space can be constructed since the Fock state describing the $k$-th time bin being occupied by $i_k$ photons is obtained via
\begin{equation}
    \ket{i_k}_k = \frac{\left[\Delta R^\dagger(t_k)\right]^{i_k}}{\sqrt{i_k!\left(\Delta t\right)^{i_k}}}\ket{\text{vac}}_k.
\end{equation}
The general state of the TLS and the photonic reservoir in the time-bin basis takes the form
\begin{multline}
    \ket{\psi(t_k)} = \sum_{i_1, \dots, i_{k-1}, i_S, i_{k}, \dots, i_N} \psi_{i_1, \dots, i_{k-1}, i_S, i_{k}, \dots, i_N} \\ \times \ket{i_1, \dots, i_{k-1}, i_S, i_k, \dots, i_N} 
\end{multline}
where $i_S \in \{g,e\}$ denotes the TLS being either in the ground ($g$) or the excited state ($e$) while the index $i_j$, $j \in \{1,\dots,N\}$, describes the occupation of the $j$-th time bin. Time is assumed to run from $t_1$ to $t_N$.
The coefficient tensor $\psi_{i_1, \dots, i_{k-1}, i_S, i_{k}, \dots, i_N}$ is, in general, $2 p^N$ dimensional where $(p-1)$ is the maximum number of photons per time bin considered. The dimension of the Hilbert space, thus, grows exponentially with the number of time bins. 
To effectively reduce the dimension of the Hilbert space and enable an efficient numerical calculation of the dynamics, the coefficient tensor is decomposed into a product of matrices via a series of singular value decompositions. The state can subsequently be written as
\begin{multline}
    \ket{\psi(t_k)} = \sum_{i_1, \dots, i_{k-1}, i_S, i_k, \dots , i_N} A^{i_1} \cdots A^{i_{k-1}} A^{i_S} A^{i_k} \cdots A^{i_N} \\ \times \ket{i_1, \dots, i_{k-1}, i_S, i_k, \dots, i_N}.
\end{multline}
This way, a time-local description is obtained since each matrix $A^{i_j}$ refers to a specific time bin while the matrix $A^{i_S}$ describes the TLS. Furthermore, the singular values provide an opportunity to quantify the entanglement between neighboring sites and allow a justified truncation of the Hilbert space. The idea of the truncation scheme is to neglect the least entangled and, thus, least important parts of the Hilbert space via a limitation of the bond dimension.

The time evolution is eventually performed by contracting and decomposing the time evolution operator, the TLS bin, and the involved time bins where a swapping algorithm allows for the efficient inclusion of the non-Markovian feedback contributions.

\subsection{Quantized pulses}
\label{QuantizedPulses}

Without a quantized pulse, that is, for a reservoir initially in the vacuum state, each of the time bins can be initialized in the vacuum state individually since the reservoir is found in a product state and there is no entanglement between the bins.

If we, however, drive the TLS with a quantized pulse, the involved reservoir bins become temporally entangled \cite{Guimond2017}. In the case of a single-photon pulse, the initial state of the reservoir is given as
\begin{equation}
    \ket{\psi(t_0)}_\text{res} = a^\dagger_f \ket{0, \dots, 0}
\end{equation}
where $a^\dagger_f$ is the creation operator of a wave packet with normalized pulse shape $f(t)$ for which
\cite{Loudon}
\begin{equation}
    a_f^\dagger = \int dt f(t) r^\dagger_t,\quad \int dt \left|f(t)\right|^2 = 1,\quad \left[a_f, a_f^\dagger \right] = 1. \label{eq:wavepacket}
\end{equation}

This formulation in the time domain can be related to the description in frequency space 
\begin{equation}
a_f^\dagger = \int \mathrm{d}\omega f(\omega) r_\omega^\dagger
\end{equation}
via the Fourier transform of the coefficients,
\begin{equation}
f(\omega) = \frac{1}{\sqrt{2\pi}}\int \mathrm{d}t f(t) e^{-i(\omega-\omega_0)t}.
\end{equation}
For reasons of clarity, here, we focus on the state of the reservoir exclusively. Typically, the reservoir and the TLS are initially separable so that the TLS can be initialized independently.
In the time-bin basis, assuming the pulse shape to be constant during one time step, that is, $f(t) = f_k$, $t \in \left[t_k, t_{k+1}\right.\!\!\left[\right.$, $k \in \{1, \dots , N\}$, the initial state is
\begin{equation}
    \ket{\psi(t_0)}_\text{res} = \sum_{k = 1}^N f_k \Delta R^\dagger(t_k)\ket{0, \dots, 0}.
\end{equation}
A rectangular pulse which starts at $t_\text{start} = t_1$ and ends at $t_\text{end} = t_2$, for example, yields
\begin{multline}
 \ket{\psi(t_0)}_\text{res} = \frac{1}{\sqrt{2 \Delta t}}\left[\Delta R^\dagger(t_1)+\Delta R^\dagger(t_2) \right] \ket{0, \dots, 0} \\= \frac{1}{\sqrt{2}}\left[ \ket{1, 0}_{1,2} + \ket{0,1}_{1,2} \right] \otimes \ket{0, \dots, 0}_{3,\dots,N}
 \label{eq:1-photon-rect-pulse}
\end{multline}
where the subscripts in the second line indicate the associated time bins.
The time bins involved in the pulse cannot be initialized separately but due to their entanglement have to be initialized collectively and are subsequently decomposed into the MPS form.
For a rectangular single-photon pulse running from time $t_\text{start} = p_1 \Delta t$ to $t_\text{end} = p_N \Delta t$, the matrices $A[p]^{i_p}$ where $p$ denotes the time step and $i_p$ is the respective physical index take the form
\begin{align}
    A[p_1]^1 = \begin{pmatrix} 1 & 0\end{pmatrix},& \quad A[p_1]^2 = \begin{pmatrix}0 & 1 \end{pmatrix}, \\
    A[p_k]^1 = \begin{pmatrix} 1 & 0 \\
    0 &\sqrt{\frac{1}{k}} \end{pmatrix},& \quad A[p_k]^2 = \begin{pmatrix} 0 & \sqrt{\frac{k-1}{k}} \\
    0 &0 \end{pmatrix}, \\
    A[p_N]^1 = \begin{pmatrix} 0 & \sqrt{\frac{N-1}{N}}\end{pmatrix}^T,& \quad A[p_N]^2 = \begin{pmatrix} \sqrt{\frac{1}{N}}& 0 \end{pmatrix}^T
\end{align}
with $1<k<N-1$.

To simulate $n$-photon pulses, we generalize the formalism accordingly to
\begin{equation}
    \ket{\psi(t_0)}_\text{res} = \frac{1}{\sqrt{n!}} \left(a_f^\dagger \right)^n \ket{0, \dots, 0}
\end{equation}
so that, for example, for $n=2$ a pulse with the same shape as considered in Eq.~\eqref{eq:1-photon-rect-pulse} results in the initial state
\begin{multline}
    \ket{\psi(t_0)}_\text{res} =  \frac{1}{\sqrt{4}} \left[\ket{2,0}_{1,2} + \ket{0,2}_{1,2} + \sqrt{2} \ket{1,1}_{1,2} \right] \\ \otimes \ket{0,\dots,0}_{3,\dots,N}.
\end{multline}

The explicitly decomposed matrices for the two- and three-photon pulses we consider in this work are given in the Appendix~\ref{Ap:1}. The extension to four and more photons is straightforward. In principle, the formalism allows the inclusion of arbitrary numbers of photons clearly surpassing the common one- or two-photon limit; however, the required computational resources increase with the number of excitations in the system since the bond dimension grows with the considered number of photons as we see in more detail in Sec.~\ref{MPS_results}.
The numerical MPS calculations in this Paper were performed using the ITensor library \cite{itensor}.

\subsection{Benchmark for the pulse inclusion}
\label{Benchmark}
As a benchmark for the interaction of the TLS with quantized pulses we consider the problem within the Heisenberg picture and assume $t<\tau$, that is, we concentrate on the time before the feedback mechanism comes into play.
Using the Hamiltonian $\mathcal{H}'$ given in Eq.~\eqref{eq:H_int}, we derive differential equations for the operators $E(t)$, $\sigma_-(t)$, and $r_\omega(t)$ which are time-dependent in the Heisenberg picture and obtain
\begin{align}
    \frac{\text{d}}{\text{d}t} E(t) &=  i \int d\omega g(\omega) \left[e^{i\left(\omega -\omega_0\right)t} r_\omega^\dagger(t) \sigma_-(t) -\text{H.c.}\right], \label{E_main}\\ 
\frac{\text{d}}{\text{d}t} \sigma_-(t) &= -i \int d \omega g(\omega) e^{-i \left(\omega-\omega_0\right)t} \left[\mathbb{1}-2E(t)\right]r_\omega(t), \label{P_main} \\
\frac{\text{d}}{\text{d}t} r_\omega(t) &= - i g(\omega) e^{i \left(\omega-\omega_0\right)t} \sigma_-(t). \label{r_omega_main}
\end{align}
The combined state of the TLS and the reservoir is of the form $\ket{j, n}$ where $j \in \{\text{g},\text{e} \}$ denotes the TLS being either in the ground (g) or the excited state (e) while $n \in \mathbb{N}$ indicates the number of photons in the reservoir initially.
In analogy to the MPS method, the $n$-photon state can be constructed via
\begin{equation}
    \ket{j,n} = \frac{1}{\sqrt{n!}}\left(a_f^\dagger\right)^n\ket{j,0}
\end{equation}
where $a_f^\dagger$ is the creation operator of a wave packet of shape $f(t)$ with the properties given in Eq.~\eqref{eq:wavepacket}.

\begin{figure}
    \includegraphics[width=\linewidth]{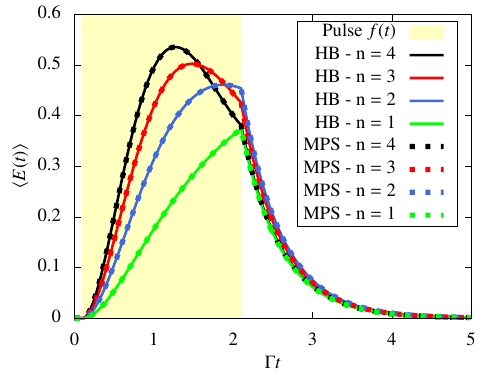}
    \caption{(Color online) Comparison of the results for the excitation of a TLS with $\Gamma \tau>5$ as a function of time obtained using the recursive Heisenberg approach (HB, solid lines) and the MPS method (dashed lines). The TLS is initially in the ground state and excited via a rectangular pulse $f(t)$ of duration $\Gamma t_p =  2$ which contains n photons, $n \in \{1, 2, 3, 4\}$.}
    \label{fig:benchmark}
\end{figure}

As for the MPS method, we assume the TLS and the reservoir to be initially separable so that we can calculate the expectation value of the occupation operator of the TLS, $\braket{E(t)} = \bra{\psi(0)}E(t)\ket{\psi(0)}$, using an initial state of the form $\ket{\psi(0)} = \ket{j,n}$.
When calculating this expectation value numerically, we find a coupling to matrix elements which for $t < \tau$ obey 
\begin{multline}
    \frac{\text{d}}{\text{d}t}\bra{i,m}E(t) \ket{k,p} = - 2 \Gamma \bra{i,m}E(t) \ket{k,p}  \\
    - \sqrt{\Gamma} \left[\sqrt{m}f^*_\tau(t) \bra{i,m-1} \sigma_{-}(t)\ket{k,p} \right.  \\
    \left. \qquad \quad+ \sqrt{p} f_\tau(t) \bra{i,m}\sigma_+(t)\ket{k,p-1} \right], \label{E_mat_main}
    \end{multline}
    \begin{multline}
    \frac{\text{d}}{\text{d}t} \bra{i,m}\sigma_{-}(t)\ket{k,p} = - \Gamma \bra{i,m}\sigma_{-}(t)\ket{k,p} \\
    - \sqrt{\Gamma}\sqrt{p}f_\tau(t)\left[ \braket{i,m|k,p-1} - 2 \bra{i,m}E(t)\ket{k,p-1}\right] \label{P_mat_main}
\end{multline}
with $f_\tau(t) = f\left(t-\frac{\tau}{2}\right) e^{i\omega_0\frac{\tau}{2}} - f\left(t+\frac{\tau}{2}\right)e^{-i\omega_0\frac{\tau}{2}}$. 
The detailed derivations can be found in the Appendix~\ref{Ap:2}. 
We solve the problem with $n$ excitations in the system by recursively inserting the results for the case of $n-1$ excitations \cite{Wang2012} and, in principle, an analytical calculation is possible.
In Fig.~\ref{fig:benchmark}, the MPS results for the dynamics of a TLS interacting with rectangular pulses of duration $\Gamma t_p = 2$ containing up to four photons are compared with those obtained using the recursive Heisenberg approach introduced in this section. Here, feedback effects do not come into play since $\Gamma \tau > 5$. The results coincide perfectly, confirming the validity of the pulse implementation in the MPS framework.

\section{Addressing the atom-photon bound state}
\label{MPS_results}

The MPS formalism introduced in Sec.~\ref{MPS_method} allows the numerically exact simulation of the non-Markovian dynamics of quantum few-level systems driven via quantized pulses containing different numbers of photons. 
For the system we focus on, a TLS inside a semi-infinite waveguide, feedback effects such as the possibility to excite a bound state in the continuum arise. This phenomenon manifests as a stabilization of the excitation probability (henceforth termed excitation for brevity) of the TLS pointing to the excitation of an atom-photon bound state \cite{Longo2010}.

Due to the implemented feedback mechanism, a signal emitted towards the mirror returns to the TLS after the delay time $\tau$ and interferes with the signal that is emitted from the TLS at that moment as illustrated in Fig.~\ref{fig:TLS_Setup}. The effect of the interference depends on the feedback phase $\varphi \equiv \omega_0 \tau$ where $\omega_0$ is the transition frequency of the TLS. If the condition $\varphi = 2 \pi m$, $m \in \mathbb{N}$, is met, the interference potentially leads to a stabilization of the excitation in the emitter and the trapping of a certain amount of excitation between the TLS and the mirror.
For a feedback phase $\varphi \neq 2\pi m$, in the long-time limit, the emitter inevitably decays to the ground state \cite{Carmele2020b,Finsterholzl2020}.
Henceforth, we assume that a feedback phase $\varphi = 2\pi m$ is implemented so that the excitation of an atom-photon bound state is, in principle, possible.
In our analysis, we concentrate on the steady-state excitation of the emitter, that is, the excitonic component of the atom-photon bound state, as a measure of the excitation trapping.

\begin{figure}
    \includegraphics[width=\linewidth]{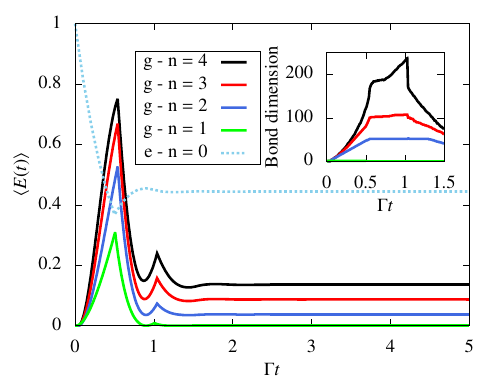}
    \caption{(Color online) Excitation of a TLS with $\Gamma\tau = 0.5$ as a function of time. The TLS is either initially excited (e) and decays in the vacuum (dashed light blue line), $n = 0$, or initially in the ground state (g) and excited via a rectangular pulse of duration $\Gamma t_p = 0.5$ which contains $n$ photons, $n \in \{1,2,3,4\}$. Inset: Bond dimension of the time bins in the MPS implementation for an $n$-photon pulse.}
    \label{fig:rect_pulses_MPS1}
\end{figure}

To begin with, we consider the regime  of medium delay times. In Fig.~\ref{fig:rect_pulses_MPS1}, the dynamics for a TLS with $\Gamma \tau = 0.5$ subjected to rectangular pulses of duration $\Gamma t_p = 0.5$ containing up to four photons is shown and compared to the case of an initially excited emitter decaying spontaneously in the vacuum.
There are different scenarios in which an atom-photon bound state is excited:
On the one hand, a stabilization of the excitation can be observed for an initially excited emitter that does not fully decay in the vacuum (dashed light blue line). In this case, the amount of excitation trapped in the system decreases monotonously with $\Gamma\tau$ \cite{Tufarelli2013}.
On the other hand, we can evoke a stabilization of the excitation in a TLS that is initially in the ground state using multi-photon pulses. In this case, the shape of the pulse and the contained number of photons additionally influence the trapped excitation. 
For a TLS initially in the ground state, a single-photon pulse does not cause a stabilization at a finite amount of excitation (solid green line, first from the bottom). We need at least two photons in the pulse to evoke such behavior where the first photon partially excites the emitter and due to the scattering of the remaining photons, a non-zero steady state can be reached \cite{Calajo2019a}. 

A stabilization can be observed for the two-photon pulse (solid blue line, second from the bottom). Further increasing the number of photons in the pulse results in an increasing steady-state excitation for the system under consideration as we see in the case of a three- (solid red line, second from the top) and a four-photon pulse (solid black line, first from the top). Here, the steady-state excitation the initially excited emitter relaxes to clearly exceeds the one that can be reached using the considered pulses with up to four photons. This is, however, not always the case as we will see below.
The specific pulse width was chosen since for the given system parameters we found it to yield the highest steady-state excitation possible with rectangular pulses. Generalizing this 
observation, we found that in the considered range of parameters the highest possible steady-state excitation was evoked by pulses of width $\Gamma t_p \lesssim \Gamma\tau$.

In the inset of Fig.~\ref{fig:rect_pulses_MPS1}, the bond dimension of the time bins in the MPS implementation for different numbers of photons is shown. The bond dimension quantifies the entanglement of the time bins and, hence, functions as a measure of the required computational resources. This way, it gives an impression of the scaling of the MPS method with the number of excitations. In the case of a single-photon pulse, the maximum is reached at $t = t_\text{start}$ where the pulse starts. For a two- or three-photon pulse, the bond dimension reaches its maximum after one feedback round trip time, that is, at $t = t_\text{start} + \tau$, before decaying again while for the four-photon pulse this is the case after two feedback intervals at $t = t_\text{start} + 2 \tau$.

\begin{figure}
    \includegraphics[width=\linewidth]{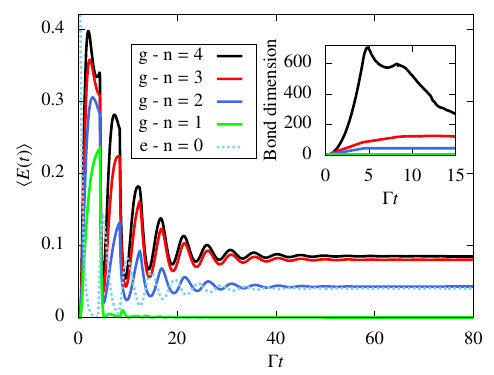}
    \caption{(Color online) Excitation of a TLS with $\Gamma\tau = 4$ as a function of time. The TLS is either initially excited (e) and decays in the vacuum (dashed light blue line), $n = 0$, or initially in the ground state (g) and excited via a rectangular pulse of duration $\Gamma t_p = 4$ which contains $n$ photons, $n \in \{1,2,3,4\}$. Inset: Bond dimension of the time bins in the MPS implementation for an $n$-photon pulse.}
    \label{fig:rect_pulses_MPS2}
\end{figure}

Next, we turn to the strongly non-Markovian regime characterized by $\Gamma \tau \gg 1$. 
The dynamics of the excitation of a TLS subjected to feedback with $\Gamma\tau = 4$ is shown in Fig.~\ref{fig:rect_pulses_MPS2}. The emitter is either initially excited and decays spontaneously in the vacuum or starts in the ground state and is excited by rectangular pulses of duration $\Gamma t_p = 4$ containing up to four photons. In this regime, the excitation of the atom-photon bound state via multi-photon pulses is more effective than in the regime of short delay times.

Since we are interested in ways to control and, in particular, maximize the trapping probability, we note that the steady-state excitation of the TLS presented in Fig.~\ref{fig:rect_pulses_MPS2} for the considered rectangular two-photon pulse approximately matches the excitation at which the initially excited TLS in the vacuum stabilizes, $\braket{E(t_\infty)} \equiv \lim_{t \rightarrow \infty} \langle E(t) \rangle = 0.040$. The rectangular three-photon pulse, by contrast, results in a steady-state excitation that clearly exceeds this value. A fourth photon in the pulse additionally increases the steady-state excitation slightly, $\braket{E(t_\infty)} = 0.085$. Comparing the two excitation schemes this corresponds to an increase of around 110\,\%.
Thus, our findings suggest that the quantum optical preparation of an excited emitter via multi-photon pulses can be significantly more effective than via an initially excited emitter in the regime of large delay times. Here, the quantum pulse induces a stimulated emission process that enhances the trapping probability in comparison to the spontaneous decay of an initially excited emitter.

In the inset of Fig.~\ref{fig:rect_pulses_MPS2}, the bond dimension of the time bins for the different numbers of excitations in the system is presented so that an assessment of the required computational resources is possible. Comparing it to the inset of Fig.~\ref{fig:rect_pulses_MPS1}, we see that in addition to the number of photons in the pulse, a long delay time is the major numerical cost factor.

\section{Conclusion and outlook}
\label{Conclusion}

We studied the interaction of a TLS with the electromagnetic field inside a semi-infinite one-dimensional waveguide within the MPS framework. In this system, multi-photon pulses can excite an atom-photon bound state.

The effectiveness of the excitation scheme depends on the system parameters. In the regime of small delay times, the excitonic component of the bound state dominates and its excitation is most effective via the spontaneous emission of an initially excited emitter. In the strongly non-Markovian regime of large delay times, our analysis for up to four photons showed that via multi-photon pulses the emitter can be stabilized at a steady-state excitation exceeding that of an initially excited TLS decaying in the vacuum significantly. For the parameters we considered, we found an increase of 110\,\%.
This shows that multi-photon pulses are a versatile tool for the excitation of the atom-photon bound state since they induce a stimulated emission process that can enhance the trapping probability in comparison to the spontaneous decay of an initially excited emitter, especially in the strongly non-Markovian regime.
Thus, the findings suggest that it is possible to realize tailored trapping scenarios using pulse engineering 
which can be an important step on the path towards the implementation of effective quantum memory.

It will be interesting to extend our model to more complex systems consisting of multiple emitters where, for example, the effect of quantum pulses on the entanglement of the emitters can be studied.

\section*{Acknowledgements}
The authors gratefully acknowledge the support of the Deutsche Forschungsgemeinschaft (DFG) through the project B1 of the SFB 910 and from the European Unions Horizon 2020 research and innovation program under the SONAR grant Agreement No.734690.

\begin{appendices}
\section{Initialization of the pulse bins}
\label{Ap:1}

If the TLS is subjected to a quantized pulse, the tensor describing the entangled state of the involved time bins has to be decomposed into matrices allowing a time local description. 

For a rectangular two-photon pulse starting at $t_\text{start} = p_1\Delta t$ and ending at  $t_\text{end}= p_N \Delta t$, the matrices $A[p]^{i_p}$ describing time bin $p$ with corresponding physical index $i_p$ take the form
\begin{align}
    A[p_1]^1 = \begin{pmatrix} 1 & 0 & 0\end{pmatrix},& \quad A[p_1]^2 = \begin{pmatrix}0 & 1 & 0  \end{pmatrix}, \notag \\
     A[p_1]^3 = &\begin{pmatrix}0 & 0 & 1 \end{pmatrix},
\end{align}
\begin{align}
    A[p_k]^1 = \begin{pmatrix} 1 & 0 & 0 \\
    0 &\sqrt{\frac{k-1}{k}} & 0   \\ 0& 0& \sqrt{\frac{(k-1)^2}{k^2}}  \end{pmatrix},& \quad A[p_k]^2 = \begin{pmatrix} 0 & \sqrt{\frac{1}{k}} & 0 \\
    0 &0 & \sqrt{\frac{2(k-1)}{k^2}} \\ 0&0&0\end{pmatrix}, \notag \\
     A[p_k]^3 = &\begin{pmatrix} 0 & 0 & \sqrt{\frac{1}{k^2}} \\
    0 &0 & 0   \\ 0& 0& 0 \end{pmatrix},
\end{align}
\begin{align}
    A[p_N]^1 = \begin{pmatrix} 0 & 0& \sqrt{\frac{(N-1)^2}{N^2}}\end{pmatrix}^T,& \quad A[p_N]^2 = \begin{pmatrix}0 & \sqrt{\frac{2(N-1)}{N^2}}& 0 \end{pmatrix}^T, \notag \\
    A[p_N]^3 = &\begin{pmatrix} \sqrt{\frac{1}{N^2}} & 0 & 0 \end{pmatrix}^T
\end{align}
where $1<k<N-1$.
If the same pulse containing three photons is considered, we find
\begin{align}
    A[p_1]^1 = \begin{pmatrix} 1 & 0 & 0 & 0 \end{pmatrix},& \quad A[p_1]^2 = \begin{pmatrix}0 & 1 & 0 & 0 \end{pmatrix}, \notag \\
     A[p_1]^3 = \begin{pmatrix}0 & 0 & 1 & 0  \end{pmatrix},& \quad A[p_1]^4 = \begin{pmatrix}0 & 0 & 0 & 1  \end{pmatrix},
\end{align}
\begin{align}
    A[p_k]^1 &= \begin{pmatrix} 1 & 0 & 0 & 0\\
    0 &\sqrt{\frac{k-1}{k}} & 0  &0   \\ 0& 0& \sqrt{\frac{(k-1)^2}{k^2}} & 0 \\ 0&0&0&\sqrt{\frac{(k-1)^3}{k^3}}  \end{pmatrix}, \notag \\
    A[p_k]^2 &= \begin{pmatrix} 0 & \sqrt{\frac{1}{k}} & 0 & 0\\
    0 &0 & \sqrt{\frac{2(k-1)}{k^2}} & 0 \\ 0&0&0&\sqrt{\frac{3(k-1)^2}{k^3}} \\ 0&0&0&0 \end{pmatrix}, \notag \\
     A[p_k]^3 &= \begin{pmatrix} 0 & 0 & \sqrt{\frac{1}{k^2}} & 0 \\
    0 &0 & 0 & \sqrt{\frac{3(k-1)}{k^3}}   \\ 0& 0& 0 & 0 \\ 0&0&0&0\end{pmatrix}, \notag \\
    A[p_k]^4 &= \begin{pmatrix} 0 & 0 & 0 & \sqrt{\frac{1}{k^3}} \\
    0 &0 & 0 & 0   \\ 0& 0& 0 & 0 \\ 0&0&0&0\end{pmatrix},
\end{align}
\begin{align}
    A[p_N]^1 = \begin{pmatrix} 0 & 0&0&\sqrt{\frac{(N-1)^3}{N^3}}\end{pmatrix}^T,& \text{  } A[p_N]^2 = \begin{pmatrix}0&0 & \sqrt{\frac{3(N-1)^2}{N^3}}& 0 \end{pmatrix}^T, \notag \\
    A[p_N]^3 = \begin{pmatrix}0& \sqrt{\frac{3(N-1)}{N^3}} & 0 & 0 \end{pmatrix}^T,& \text{  } A[p_N]^4 = \begin{pmatrix} \sqrt{\frac{1}{N^3}}&0 & 0 & 0 \end{pmatrix}^T.
\end{align}

\section{Benchmark}
\label{Ap:2}

In order to benchmark the interaction of the TLS with quantized pulses before feedback effects set in, that is, for $t<\tau$, we derive differential equations for the operators $E(t)$, $\sigma_-(t)$, and $r_\omega(t)$ in the Heisenberg picture . 
An arbitrary Heisenberg operator $A(t)$ is related to its counterpart in the Schr\"odinger picture $A_\text{S} = A(0)$ via
\begin{equation}
    A(t) = U^\dagger(t,0) A_\text{S} U(t,0)
\end{equation}
where $U(t,0)$ is the time-evolution operator from time $0$ to time $t$.
Assuming no explicit time dependence, $A(t)$ obeys the Heisenberg equation of motion
\begin{equation}
    \frac{\text{d}}{\text{d}t}A(t) = \frac{i}{\hbar}  U^\dagger(t,0) \left[\mathcal{H}', A_S \right] U(t,0).
\end{equation}
Using the Hamiltonian $\mathcal{H}'$ given in Eq.~\eqref{eq:H_int} of the main text, we obtain 
\begin{align}
    \frac{\text{d}}{\text{d}t} E(t) &=  i \int d\omega g(\omega) \left[e^{i\left(\omega -\omega_0\right)t} r_\omega^\dagger(t) \sigma_-(t) -\text{H.c.}\right], \label{E}\\ 
\frac{\text{d}}{\text{d}t} \sigma_-(t) &= -i \int d \omega g(\omega) e^{-i \left(\omega-\omega_0\right)t} \left[\mathbb{1}-2E(t)\right]r_\omega(t), \label{P} \\
\frac{\text{d}}{\text{d}t} r_\omega(t) &= - i g(\omega) e^{i \left(\omega-\omega_0\right)t} \sigma_-(t). \label{r_omega}
\end{align}
We integrate out the reservoir modes by formally integrating Eq. \eqref{r_omega} and plugging the result
\begin{equation}
    r_\omega(t) = r_\omega(0) -i g(\omega) \int_0^t dt' e^{i\left(\omega-\omega_0\right)t'} \sigma_-(t')
\end{equation}
into Eqs.~\eqref{E} and \eqref{P}.
In analogy to the MPS method presented in Sec.~\ref{MPS_method}, we introduce the quantum noise operators $r_t^\dagger$(0) as the conjugate operators of $r_\omega^\dagger(0)$ [see Eq.~\eqref{eq:noise_op} of the main text] which allow the description of a fully quantized input pulse \cite{Gardiner1985, Gardiner2004a}.
The Markovian case where we assume a constant coupling strength between the emitter and the reservoir has been discussed extensively in the literature~\cite{Domokos2002,Konyk2016, Kiilerich2019,Liao2020}.
If a feedback mechanism at delay time $\tau > 0$ is implemented, we assume a sinusoidal frequency dependence of the coupling strength, $g(\omega) = g_0 \sin(\omega \tau /2)$, and define the delayed input operator
\begin{equation}
    r^\dagger_{t,\tau} \equiv r^\dagger_{t-\frac{\tau}{2}}(0)e^{-i \omega_0 \frac{\tau}{2}}-r^\dagger_{t+\frac{\tau}{2}}(0)e^{i\omega_0 \frac{\tau}{2}}.
\end{equation}
This way, Eqs.~\eqref{E} and \eqref{P} yield the delay differential equations \cite{Scholl2016} 
\begin{align}
   \frac{\text{d}}{\text{d}t} E(t) &= -2 \Gamma E(t) 
     - \sqrt{\Gamma} \left[r_{t,\tau}^\dagger \sigma_-(t)+\text{H.c.}\right] \notag\\
    & + \Gamma \left[ e^{-i\omega_0 \tau} \sigma_+(t-\tau)\sigma_-(t) + \text{H.c.}\right] \Theta(t-\tau), \label{eq:E_op}\\
    \frac{\text{d}}{\text{d}t} \sigma_-(t) &= - \Gamma \sigma_-(t)  - \sqrt{\Gamma}\left[ \mathbb{1} - 2 E(t) \right] r_{t,\tau} \notag \\
    &+\Gamma e^{i \omega_0 \tau}\left[\sigma_-(t-\tau) - 2 E(t) \sigma_-(t-\tau) \right] \Theta(t-\tau) \label{eq:P_op}
\end{align}
where we again used the definition of the decay rate $\Gamma = \pi g_0^2/2$ and find that after the non-negligible delay time $\tau$, feedback effects influence the dynamics. 
Since our aim is to benchmark the MPS results before feedback effects set in, we omit the time-delayed terms in Eqs.~\eqref{eq:E_op} and \eqref{eq:P_op} and, this way, avoid having to deal with two-time correlations.

We are interested in the dynamics of the expectation value of the occupation operator of the TLS, $\braket{E(t)} = \bra{\psi(0)}E(t)\ket{\psi(0)}$. Assuming the TLS and the reservoir to be initially separable, the initial state of the system can be written as $\ket{\psi(0)} = \ket{j,n}$ for a TLS initially in either the ground ($j = g$) or the excited state ($j=e$) and $n$ photons in the reservoir. This $n$-photon state can be obtained from the vacuum state of the reservoir via the creation operator $a_f^\dagger$ which describes the creation of a wave packet of shape $f(t)$ according to
\begin{equation}
    \ket{j,n} = \frac{1}{\sqrt{n!}}\left(a_f^\dagger\right)^n\ket{j,0}
\end{equation}
 where the pulse shape $f(t)$ has the properties given in Eq.~\eqref{eq:wavepacket} of the main text. Conversely, the annihilation of a photon can be described as
\begin{equation}
    r_{t}\ket{j,n} =\begin{cases}
    0, & n=0 \\
    \sqrt{n}f(t)\ket{j,n-1}, & n>0
    \end{cases}.
\end{equation}
We calculate the above expectation value using Eq.~\eqref{eq:E_op} by which it is coupled to matrix elements of the form $\bra{i,m}E(t) \ket{k,p}$ and $\bra{i,m}\sigma_{-}(t)\ket{k,p}$. Assuming $t<\tau$, these matrix elements can be obtained via
\begin{multline}
    \frac{\text{d}}{\text{d}t}\bra{i,m}E(t) \ket{k,p} = - 2 \Gamma \bra{i,m}E(t) \ket{k,p}  \\
    - \sqrt{\Gamma} \left[\sqrt{m}f^*_\tau(t) \bra{i,m-1} \sigma_{-}(t)\ket{k,p} \right.  \\
    \left. \qquad \quad+ \sqrt{p} f_\tau(t) \bra{i,m}\sigma_+(t)\ket{k,p-1} \right], \label{E_mat}
    \end{multline}
    \begin{multline}
    \frac{\text{d}}{\text{d}t} \bra{i,m}\sigma_{-}(t)\ket{k,p} = - \Gamma \bra{i,m}\sigma_{-}(t)\ket{k,p} \\
    - \sqrt{\Gamma}\sqrt{p}f_\tau(t)\left[ \braket{i,m|k,p-1} - 2 \bra{i,m}E(t)\ket{k,p-1}\right] \label{P_mat}
\end{multline}
where $f_\tau(t) = f\left(t-\frac{\tau}{2}\right) e^{i\omega_0\frac{\tau}{2}} - f\left(t+\frac{\tau}{2}\right)e^{-i\omega_0\frac{\tau}{2}}$.

\end{appendices}

\bibliographystyle{apsrev4-2}
\bibliography{MyCollection}
\clearpage

\end{document}